\documentclass[aps,prl,reprint,twocolumn]{revtex4-1}

\usepackage[draft]{todonotes}
\usepackage{siunitx}
\usepackage{xspace}
\newcommand{\MAEho}{1.51 mHa\xspace}

\newcommand{\MAEiw}{5.04 mHa\xspace}

\newcommand{\MAEig}{2.70 mHa\xspace}

\newcommand{\MAErnd}{1.49 mHa\xspace}

\newcommand{\MAErndcrig}{2.94 mHa\xspace}

\newcommand{\MAErndke}{2.98 mHa\xspace}

\newcommand{\MAEngexcited}{10.93 mHa\xspace}

\newcommand{\figref}[1]{Fig. \ref{#1}}
\usepackage{amsmath}

\newcommand{\wcexclude}[1]{#1}

\newcommand{\exclude}[1]{}

\newcommand{\SHO}{simple harmonic oscillator\xspace}
\newcommand{\DIG}{double-well inverted Gaussian\xspace}
\newcommand{\IW}{infinite well\xspace}
\newcommand{\RND}{random\xspace}
\newcommand{\DNN}{deep neural network\xspace}
\newcommand{\CNN}{convolutional neural network\xspace}
\newcommand{\ANN}{artificial neural network\xspace}
\newcommand{\PDE}{partial differential equation\xspace}
\newcommand{\MAE}{MAE\xspace}

\newcommand{\DNNs}{deep neural networks\xspace}
\newcommand{\CNNs}{convolutional neural networks\xspace}
\newcommand{\ANNs}{artificial neural networks\xspace}
\newcommand{\PDEs}{partial differential equations\xspace}

\newcommand{\MXmae}{5.90 mHa\xspace}

\begin{document}


\title{Deep learning and the Schr\"odinger equation}


\author{K. Mills}
\email[]{kyle.mills@uoit.net}
\affiliation{University of Ontario Institute of Technology}

\author{M. Spanner}
\affiliation{National Research Council of Canada}

\author{I. Tamblyn}
\email[]{isaac.tamblyn@nrc.ca}
\affiliation{University of Ontario Institute of Technology \& National Research Council of Canada}


%



\date{\today}

\begin{abstract}
\wcexclude{
We have trained a deep (convolutional) neural network to predict the ground-state energy of an electron in four classes of confining two-dimensional electrostatic potentials.  On randomly generated potentials, for which there is no analytic form for either the potential or the ground-state energy, the model was able to predict the ground-state energy to within chemical accuracy, with a median absolute error of \MAErnd.  We also investigate the performance of the model in predicting other quantities such as the kinetic energy and the first excited-state energy. 
}
\end{abstract}


\wcexclude{
   \maketitle
}

\section{Introduction}
Solving the electronic structure problem for molecules, materials, and interfaces is of fundamental importance to a large number of disciplines including physics, chemistry, and materials science. 
Since the early development of quantum mechanics, it has been noted, by Dirac among others, that ``...approximate, practical methods of applying quantum mechanics should be developed, which can lead to an explanation of the main features of complex atomic systems without too much computation" \cite{dirac}. Historically, this has meant invoking approximate forms of the underlying interactions (e.g. mean field, tight binding,
etc.), or relying on phenomenological fits to a limited number of either experimental observations or theoretical results (e.g. force fields)  \cite{Cherukara2016,Riera2016,Jaramillo-Botero2014,VanBeest1990,Ponder2003,Hornak2006,cole2007}.  The development of feature-based models is not new in the scientific literature. Indeed, prior even to the acceptance of the atomic hypothesis, van der Waals argued for an equation of state based on two physical features \cite{Waals1873}.  Machine learning (i.e. fitting parameters within a model) has been used in physics and chemistry since the dawn of the computer age.  The term machine learning is new; the approach is not.

More recently, high-level \textit{ab initio} calculations have been used to train \ANNs to fit high-dimensional interaction models \cite{Li2013,Behler2007,Morawietz2013,Behler2008,Dolgirev2016,Artrith2016}, and to make informed predictions about material properties \cite{Tian2017,Rupp2012}.  These approaches have proven to be quite powerful, yielding models trained for specific atomic species or based upon hand-selected geometric features \cite{Faber2017,Montavon2013,Lopez-Bezanilla2014}.  Hand-selected features are arguably a significant limitation of such approaches, with the outcomes dependent upon the choice of input representation and the inclusion of all relevant features.  This limitation is well known in the  fields of handwriting recognition and image classification, where the performance of the traditional hand-selected feature approach has stagnated \cite{Jia2014}.

\begin{figure*}
 \includegraphics[width=0.95\textwidth]{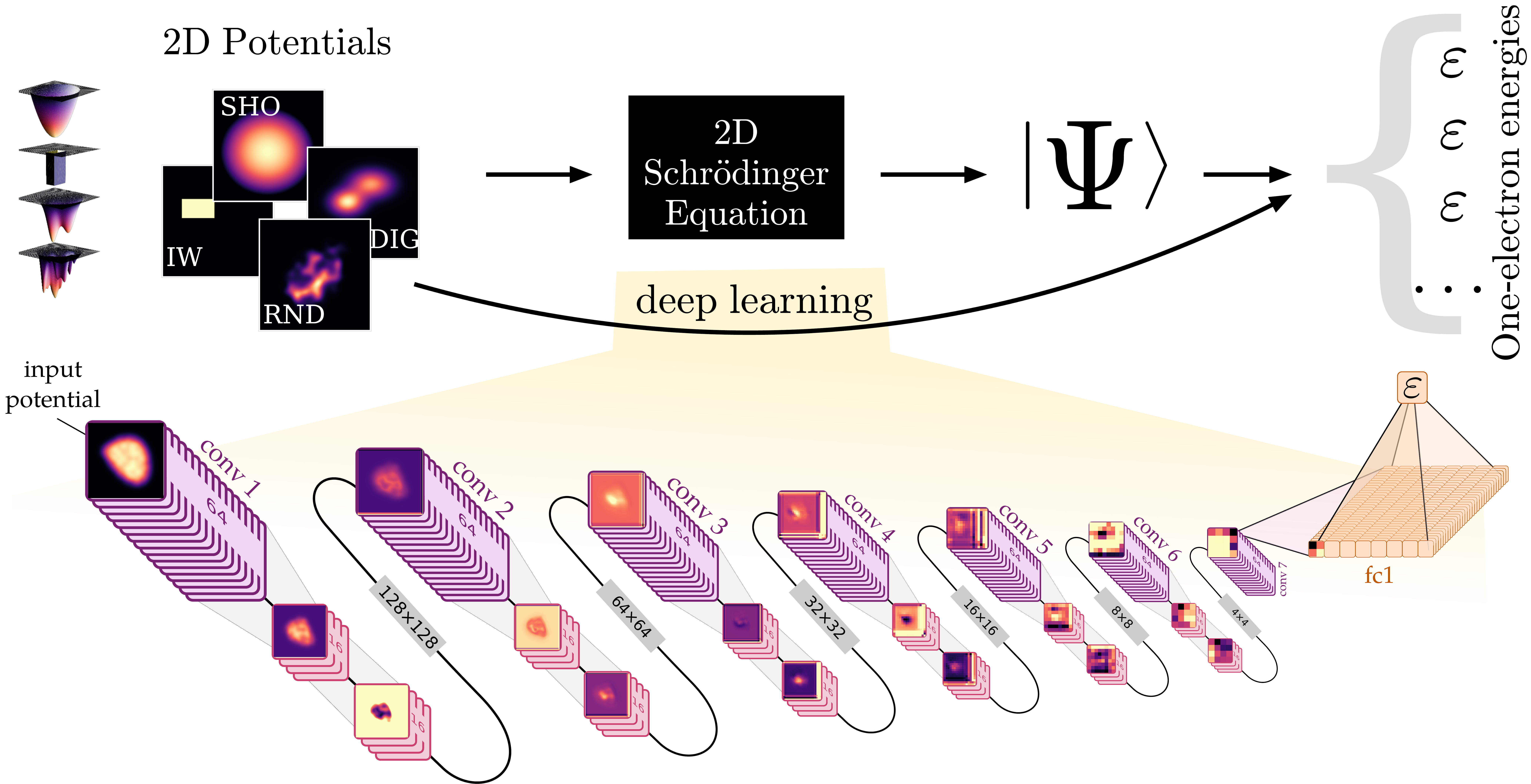}
 \caption{In this work, we use the machinery of deep learning to learn the mapping between potential and energy, bypassing the need to numerically solve the Schr\"odinger equation, and the need for computing wavefunctions.  The architecture we used (shown here) consisted primarily of convolutional layers capable of extracting relevant features of the input potentials. Two fully-connected layers at the end serve as a decision layer, mapping the automatically extracted features to the desired output quantity.  No manual feature-selection is necessary; this is a ``featureless-learning" approach.\label{schematic}}
\end{figure*}

Such feature-based approaches are also being used in materials discovery \cite{Curtarolo2003,Hautier2010,Saad2012} to assist materials scientists in efficiently targeting their search at promising material candidates. Unsupervised learning techniques have been used to identify phases in many-body atomic configurations \cite{Wang2016}. In previous work, an \ANN was shown to interpolate the mapping of position to wavefunction for a specific electrostatic potential \cite{Monterola2001,Shirvany2008, Mirzaei2010}, but the fit was not transferable, a limitation also present in other applications of \ANNs to \PDEs \cite{VanMilligen1995,Carleo2016}. By transferable, we mean that a model trained on a particular form of \PDE will accurately and reliably predict results for examples of the same form (in our case, different confining potentials).

Machine learning can also be used to accelerate or bypass some of the heavy machinery of the \textit{ab initio} method itself. In \cite{Snyder2012}, the authors replaced the kinetic energy functional within density functional theory with a machine-learned one, and in \cite{Brockherde2016} and \cite{Yao2016}, the authors ``learned'' the mappings from potential to electron density, and charge density to kinetic energy, respectively.

Here, we use a fundamentally different approach inspired by the successful application of deep convolutional neural networks to problems in computer vision \cite{Lecun1998,Simard2003,ciresan2011flexibles,Szegedy2014} and computational games \cite{Silver2016,Mnih2013}. Rather than seeking an appropriate input representation to capture the relevant physical attributes of a system, we train a highly flexible model on an enormous collection of ground-truth examples.  In doing so, the deep neural network learns \textit{both the features (in weight space) and the mapping} required to produce the desired output.  This approach does not depend on the appropriate selection of input representations and features; we provide the same data to both the deep neural network and the numerical method. As such, we call this ``featureless learning''. Such an approach may offer a more scalable and parallizable approach to large-scale electronic structure problems than existing methods can offer.



In this Letter, we demonstrate the success of a \textit{featureless} machine learning approach, a convolutional \DNN, at learning the mapping between a confining electrostatic potential and quantities such as the ground state energy, kinetic energy, and first excited-state of a bound electron. The excellent performance of our model suggests deep learning as an important new direction for treating multi-electron systems in materials.

It is known that a sufficiently large artificial neural network can approximate any continuous mapping \cite{Funahashi1989,Castro2000} but the cost of optimizing such a network can be prohibitive.  Convolutional neural networks make computation feasible by exploiting the spatial structure of input data \cite{Krizhevsky2012}, similar to how the neurons in the visual cortex function \cite{Hubel1968}.  When multiple convolutional layers are included, the network is called a deep convolutional neural network, forming a hierarchy of feature detection \cite{Bengio:2009}.  This makes them particularly well suited to data rooted in physical origin \cite{Mehta2014,Lin2016}, since many physical systems also display a structural hierarchy.  Applications of such a network structure in the field of electronic structure, however, are few (although recent work focused on training against a geometric matrix representation looks particularly promising \cite{Schutt2017}).

\section{Methods}
\subsection{Training set: choice of potentials}

Developing a deep learning model involves both the design of the network architecture and the acquisition of training data.  The latter is the most important aspect of a machine learning model, as it defines the transferability of the resulting model.  We investigated four classes of potentials: simple harmonic oscillators (SHO), ``infinite" wells (IW, i.e. ``particle in a box''), double-well inverted Gaussians (DIG), and random potentials (RND).  Each potential can be thought of as a grayscale image: a grid of floating-point numbers. 

\subsection{Numerical solver}

\begin{figure}
 \includegraphics[width=0.85\columnwidth]{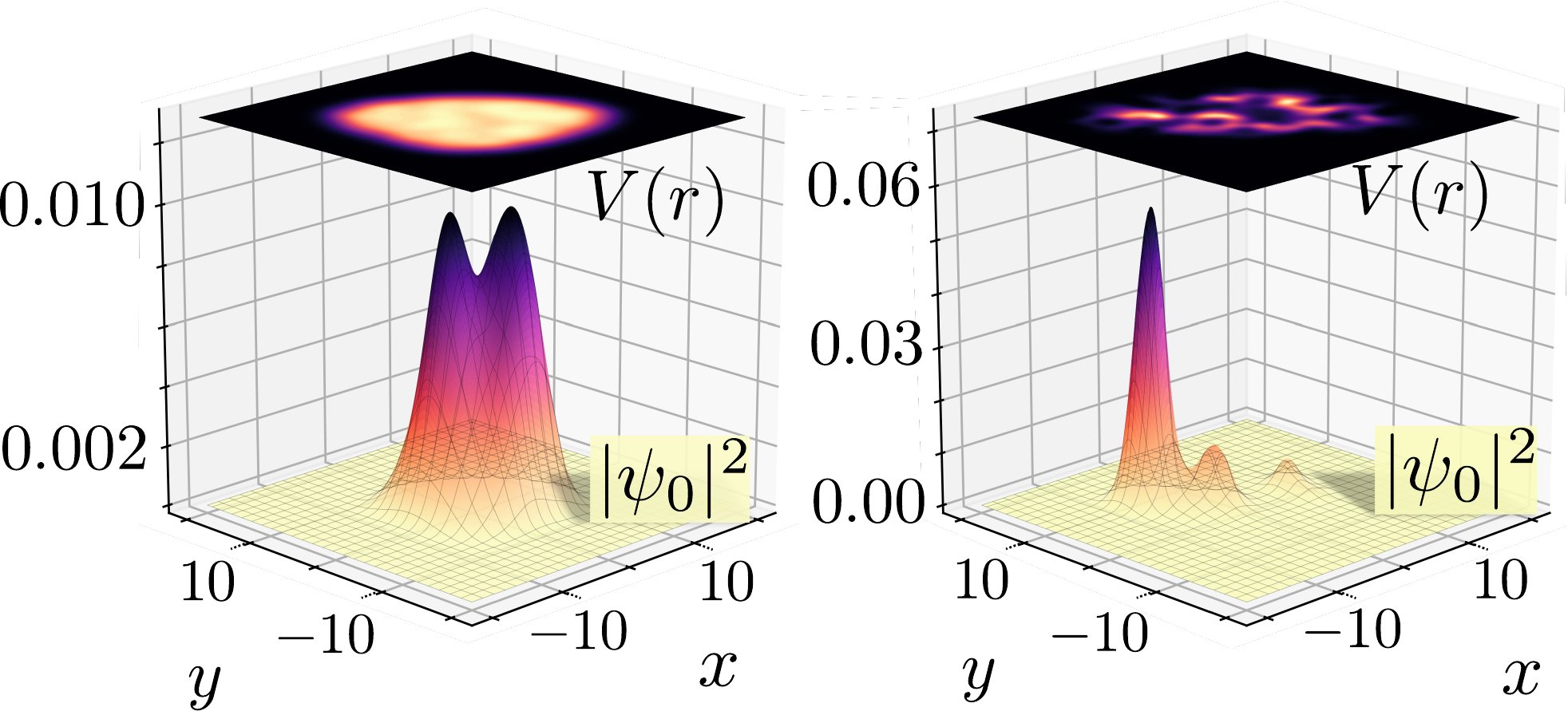}
 \caption{Wavefunctions (probability density) $|\psi_0|^2$ and the corresponding potentials $V(r)$ for two \RND potentials. \label{wavefunctions}}
\end{figure}

We implemented a standard finite-difference \cite{Press2007} method to solve the eigenvalue problem 
\wcexclude{
\begin{equation}
\hat H\psi\equiv (\hat T + \hat V)\psi  = \varepsilon\psi
\end{equation}
}
for each potential $V$ we created.  The potentials were generated with a dynamic range and length scale suitable to produce ground-state energies within a physically relevant range.  With the \RND potentials, special care was taken to ensure that some training examples produced non-trivial wavefunctions (\figref{wavefunctions}). Atomic units are used, such that $\hbar = m_\mathrm{e} = 1$.  The potentials are represented on a square domain from $-20$ to $20$ a.u., discretized on a $256\times 256$ grid.  As the \SHO potentials have an analytic solution, we used this as reference with which to validate the accuracy of the solver. The median absolute error between the analytic and the calculated energies for all \SHO potentials was $0.12$ mHa.
We discuss the generation of all potentials further in the Appendices.

The \SHO presents the simplest case for a \CNN as there is an analytic solution dependent on two simple parameters ($k_x$ and $k_y$) which uniquely define the ground-state energy of a single electron ($\varepsilon_0=\frac{\hbar}{2}(\sqrt{k_x} + \sqrt{k_y})$).  Furthermore, these parameters represent a very physical and visible quantity: the curvature of the potential in the two primary axes.  Although these parameters are not provided to the neural network explicitly, the fact that a simple mapping exists means that the \CNN need only learn it to accurately predict energies.

A similar situation exists for the \IW.  Like the \SHO, the ground state energy depends only on the width of the well in the two dimensions ($\varepsilon_0 = \frac{1}{2}\pi^2 \hbar^2 (L_x^{-2} +L_y^{-2}) $).  It would be no surprise if even a modest network architecture is able to accurately ``discover" this mapping.  An untrained human, given a ruler, sufficient examples, and an abundance of time would likely succeed in determining this mapping.

The \DIG dataset is more complex in two respects. First, the potential, generated by summing a pair of 2D-Gaussians, depends on significantly more parameters; the depth, width, and aspect ratio of each Gaussian, as well as the relative positions of the wells will impact the ground state energy.  Furthermore, there is no known analytical solution for a single electron in a potential well of this nature.  There is, however, still a concise function which describes the underlying potential, and while this is not directly accessible to the \CNN, one must wonder if the existence of such simplifies the task of the \CNN.  Gaussian confining potentials appear in works relating to quantum dots \cite{Gharaati2010,Gomez2008}.

The \RND dataset presents the ultimate challenge.  Each \RND potential is generated by a multi-step process with randomness introduced at numerous steps along the way.  There is no closed-form equation to represent the potentials, and certainly not the eigenenergies.  A \CNN tasked with learning the solution to the Schr\"odinger equation through these examples would have to base its predictions on many individual features, truly ``learning'' the mapping of potential to energy.  One might question our omission of the Coulomb potential as an additional canonical example.  The singular nature of the Coulomb potential is difficult to represent within a finite dynamic range, and, more importantly, the electronic structure methods that we would ultimately seek to reproduce already have frameworks in place to deal with these singularities (e.g. pseudopotentials).

\subsection{Deep neural network}

\begin{figure}
 \includegraphics[width=0.9\columnwidth]{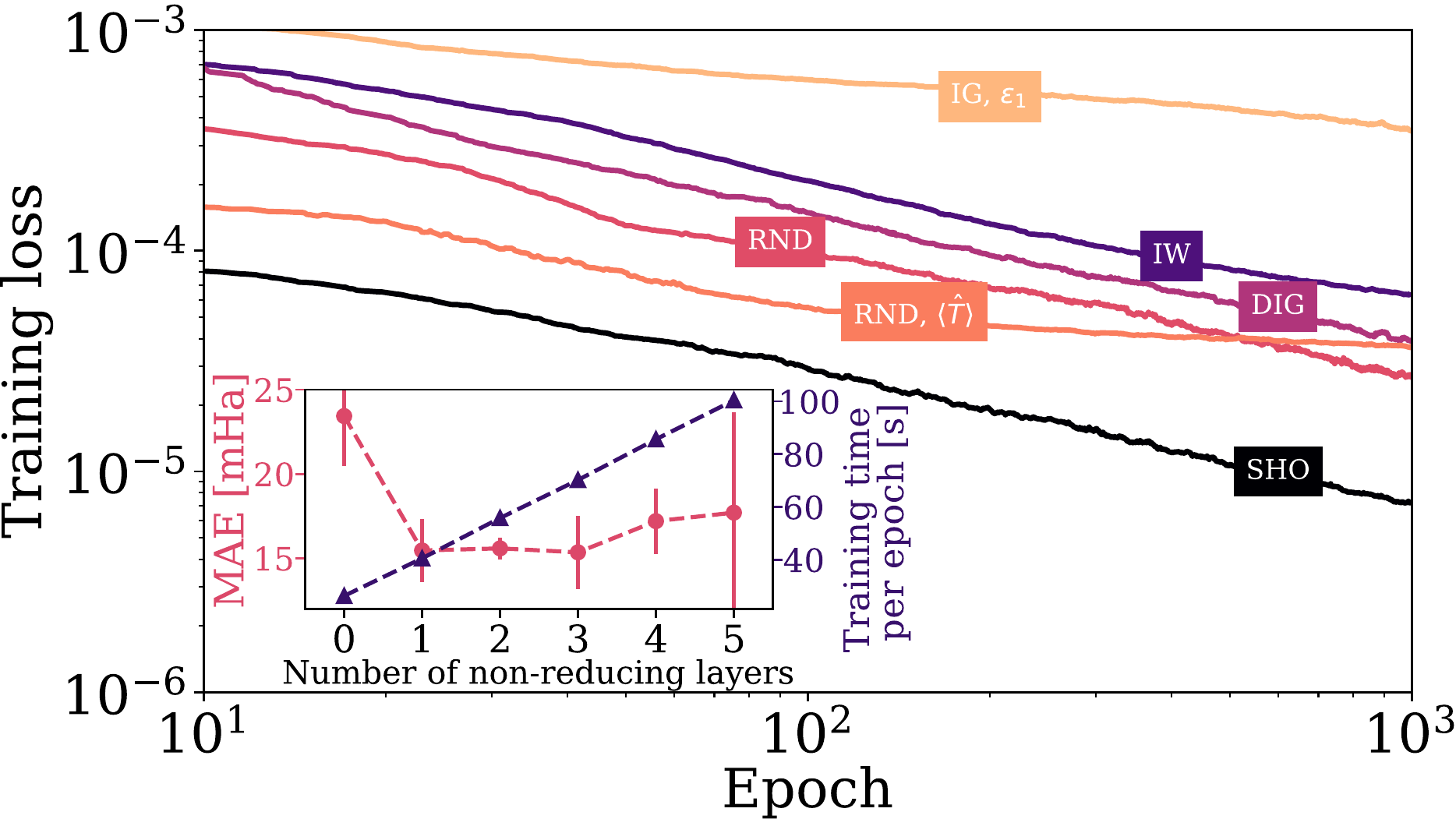}
 \caption{The training loss curve for each model we trained.  
 Since the training loss is based upon the training datasets, it does not necessarily indicate how well the model generalizes to new examples. The convergence seen here indicates that 1000 epochs is an adequate stopping point; further training would produce further reduction in loss, however 1000 epochs provides sufficient evidence that the method performs well on the most interesting (i.e. random) potentials.  In the inset, we see that two non-reducing convolution layers is a consistent balance of training time and low error. \label{lossdecrease}}
\end{figure}


We chose to use a simple, yet deep neural network architecture (shown in \figref{schematic}) composed of a number of repeated units of convolutional layers, with sizes chosen for a balance of speed and accuracy (inset of \figref{lossdecrease}). 

We use two different types of convolutional layers, which we call ``reducing'' and ``non-reducing''.

The 7 reducing layers operate with filter (kernel) sizes of $3\times 3$ pixels.  Each reducing layer operates with 64 filters and a stride of $2\times 2$, effectively reducing the image resolution by a factor of two at each step.  In between each pair of these reducing convolutional layers, we have inserted two convolutional layers (for a total of 12) which operate with $16$ filters of size $4\times4$. These filters have unit stride, and therefore preserve the resolution of the image.  The purpose of these layers is to add additional trainable parameters to the network.  All convolutional layers have ReLU activation.

The final convolutional layer is fed into a fully-connected layer of width 1024, also with ReLU activation. This layer feeds into a final fully-connected layer with a single output.  This output is the output value of the DNN.  It is used to compute the mean-squared error between the true label and the predicted label, also known as the loss.

We used the AdaDelta \cite{Zeiler2012} optimization scheme with a global learning rate of 0.001 to minimize this loss function (\figref{lossdecrease}), monitoring its value as training proceeded.  We found that after 1000 epochs (1000 times through all the training examples), the loss no longer decreased significantly.

We built a custom TensorFlow \cite{GoogleResearch2015} implementation in order to make use of 4 graphical processing units (GPUs) in parallel.  We placed a complete copy of the neural network on each of the 4 GPUs, so that each can compute a forward and back-propagation iteration on one full batch of images.  Thus our effective batch size was 1000 images per iteration (250 per GPU).  After each iteration, the GPUs share their independently computed gradients with the optimizer and the optimizer moves the parameters in the direction that minimizes the loss function.  Unless otherwise specified, all training datasets consisted of 200,000 training examples and training was run for 1000 epochs.  All reported errors are based on evaluating the trained model on validation datasets consisting of 50,000 potentials not accessible to the network during the training process.


\section{Results}
\begin{figure*}
 \includegraphics[width=0.95\textwidth]{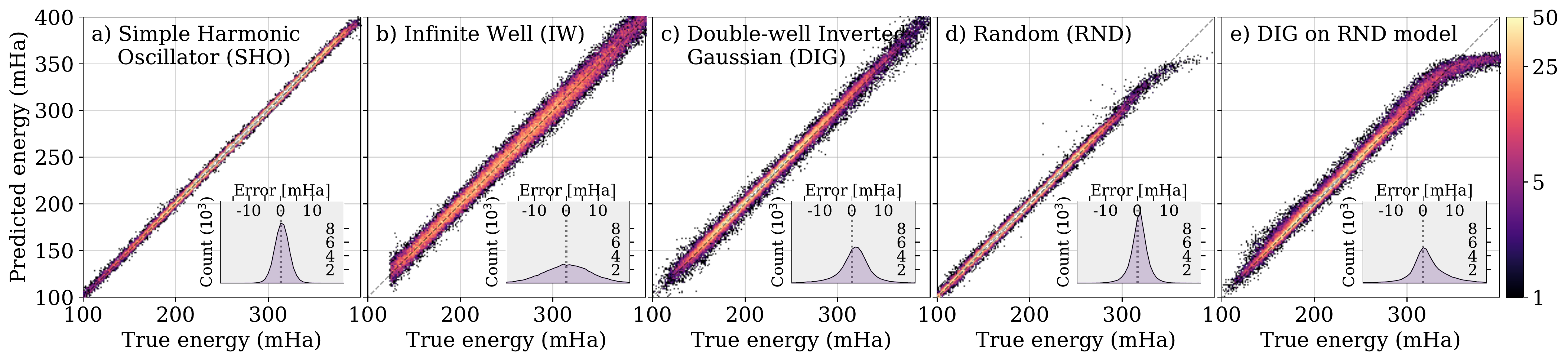}
 \caption{Histograms of the true vs. predicted energies for each example in the test set indicate the performance of the various models.  The insets show the distribution of error away from the diagonal line representing perfect predictions. A $1\ \text{mHa}^2$ square bin is used for the main histograms, and a 1 mHa bin size for the inset histogram. During training, the neural network was not exposed to the examples on which theses plots are based. The higher error at high energies in (d) is due to fewer training examples being present the dataset at these energies. The histogram shown in (d) is for the further-trained model, described in the text. \label{results01}}
\end{figure*}

\figref{results01}(a-d) displays the results for the \SHO, \IW, \DIG, and \RND potentials.  The \SHO, being one of the simplest potentials, performed extremely well. The trained model was able to predict the ground state energies with a median absolute error (\MAE) of \MAEho.

The \IW potentials performed moderately well with a \MAE of \MAEiw. This is notably poorer than the \SHO potentials, despite their similarity in being analytically dependent upon two simple parameters. This is likely due to the sharp discontinuity associated with the \IW potentials, combined with the sparsity of information present in the binary-valued potentials.

The model trained on the \DIG potentials performed moderately well with a \MAE of \MAEig and the \RND potentials performed quite well with a \MAE of 2.13 mHa.  We noticed, however, that the loss was not completely converged at 1000 epochs, so we provided an additional 200,000 training examples to the network and allowed it to train for an additional 1000 epochs.  With this added training, the the model performed exceptionally well, with a \MAE of \MAErnd, below the threshold of chemical accuracy (1 kcal/mol, 1.6 mHa).  In \figref{results01}(d), it is evident that the model performs more poorly at high energies, a result of the relative absence of high-energy training examples in the dataset. Given the great diversity in this latter set of potentials, it is impressive that the \CNN was able to learn how to predict the energy with such a high degree of accuracy.

\begin{figure}
 \includegraphics[width=0.9\columnwidth]{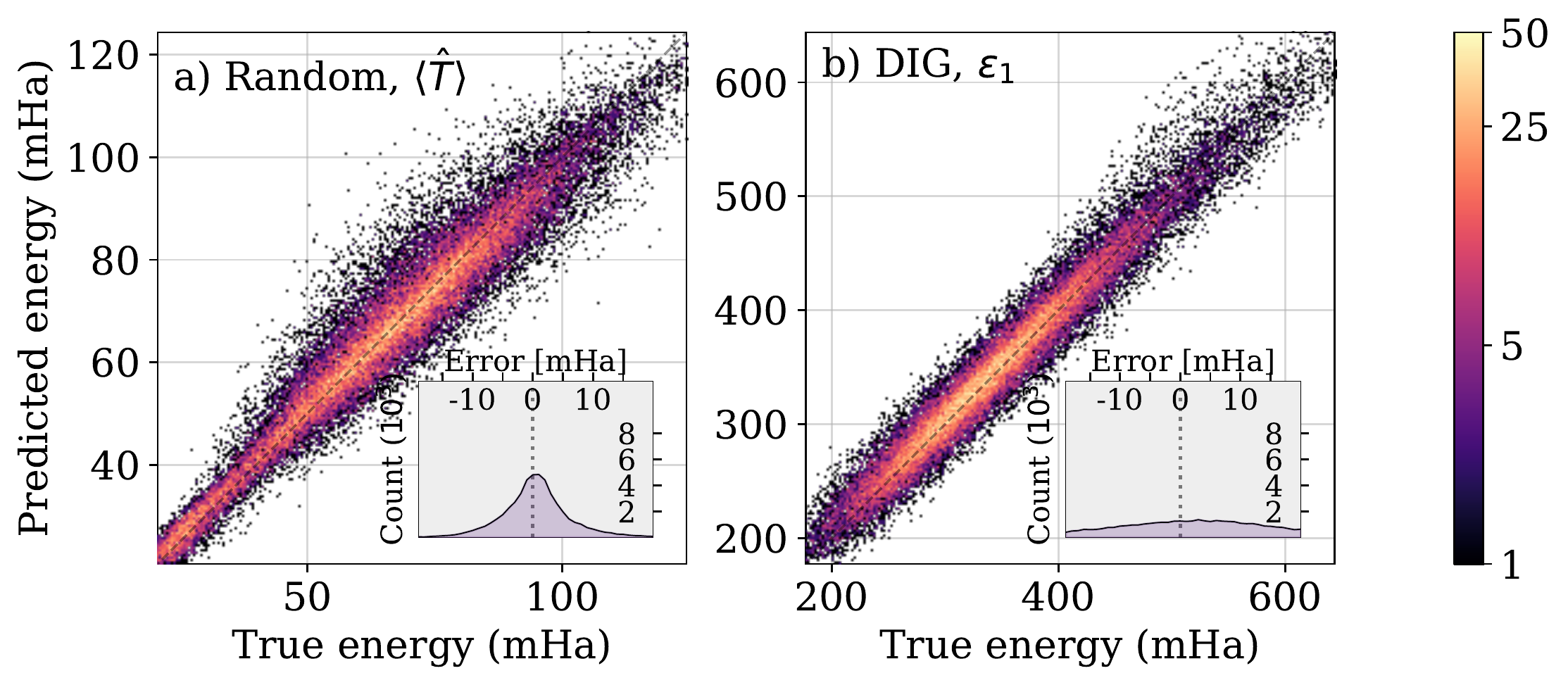}
 \caption{Histograms of the true vs. predicted energies for the model trained on the (a) kinetic energy, and (b) excited-state energy of the \DIG. \label{results02}}
\end{figure}
\exclude{
 \begin{table}
 \caption{\label{resultsTable}}
 \begin{ruledtabular}
 \begin{tabular}{lcc}
 Model & Train Examples & \MAE \\
 \hline
\SHO  							& 200k		& 	\\MAEho 				 \\ 
\IW 								& 200k		& 	\\MAEiw 				 \\ 
\DIG 							& 200k		& 	\\MAEig 				 \\ 
\RND 							& 200k		& 	\\MAErnd 			 \\ 
\RND 							& 1M 		& 	\\MAErndcrsho 		 \\ 
\RND, eval. on \DIG 				& 1M 		& 	\\MAEngexcited		 \\ 
\RND, $\langle\hat T \rangle$ 	& 1M 		&	\\MAErndke			 \\ 
 \end{tabular}
 \end{ruledtabular}
 \end{table}1
 }

Now that we have a trained model that performs well on the \RND test set, we investigated its transferability to another class of potentials.  The model trained on the \RND dataset is able to predict the ground-state energy of the \DIG potentials with a \MAE of \MAErndcrig. We can see in \figref{results02}(c) that the model fails at high energies, an expected result given that the model was not exposed to many examples in this energy regime during training on the overall lower-energy \RND dataset.  This moderately good performance is not entirely surprising; the production of the \RND potentials includes an element of Gaussian blurring, so the neural network would have been exposed to features similar to what it would see in the \DIG dataset.  However, this moderate performance is testament to the transferability of \CNN models.  Furthermore, we trained a model on an equal mixture of all four classes of potentials. It performs moderately with a MAE of \MXmae. This error could be reduced through further tuning of the network architecture allowing it to better capture the higher variation in the dataset.

The total energy is just one of the many quantities associated with these one-electron systems. To demonstrate the applicability of \DNN to other quantities, we trained a model on the first excited-state energy $\varepsilon_1$ of the \DIG potentials.  The model achieved a \MAE of \MAEngexcited. We now have two models capable of predicting the ground-state, and first excited-state energies separately, demonstrating that a neural network can learn quantities other than the ground-state energy.

The ground-state and first excited-state are both eigenvalues of the Hamiltonian.  Therefore, we investigated the training of a model on the expectation value of the kinetic energy, $\langle \hat T \rangle = \langle \psi_0 | \hat T | \psi_0 \rangle $, under the ground state wavefunction $\psi_0$ that we computed numerically for the \RND potentials. Since $\hat H$ and $\hat T$ do not 
commute, the prediction of $\langle \hat T \rangle$ can no longer be summarized as an eigenvalue problem. The trained model predicts the kinetic energy value with a \MAE of \MAErndke. While the spread of testing examples in \figref{results02}(a) suggests the model performs more poorly, the absolute error is still small.

\section{Conclusions}

\exclude{

Given the wide variety of machine learning methods which exist, one might question the choice of \CNNs over other ``simpler'' approaches. Indeed, it is only in recent years that the widespread availability of accelerator hardware and software such as GPUs has made it possible to use this computationally heavy approach for practical problems. Convolutional neural networks seem like a good choice for a number of important reasons:

\begin{enumerate}
\item Convolutional operations make use of the spatial structure of the data. Physical quantities and phenomena (e.g. the electrostatic potential, electron density, and N-body wavefunction) are amenable to \CNN.
\item The computational requirements for \CNN are easily parallelized.  The training workload for more involved problems can be distributed across large computing platforms \cite{GoogleResearch2015}.
\item The recent successes of \CNN mean there continues to be a strong community drive to develop efficient and scalable implementations. This has resulted in rapid development of scalable implementations which make use of modern hardware architectures (including multi-GPU, multi-node distributed systems). 
\item The diverse options in \DNN architectures allow for a flexible ``basis'' for more complicated problems. Other machine learning methods require feature selection. \DNN have consistently proven their ability to extract meaningful relationships from structured data without extensive intervention.
\end{enumerate}
}

We note that many other machine learning algorithms exist and have traditionally seen great success, such as kernel ridge regression \cite{Brockherde2016,Arsenault2014,Li2016,Suzuki2016,Faber2017,Lopez-Bezanilla2014} and random forests \cite{Faber2017,Ward2015}.  Like these algorithms, convolutional deep neural networks have the ability to ``learn'' relevant features and form a non-linear input-to-output mapping without prior formulation of an input representation \cite{Kearnes2016,Schutt2017}.  In our tests, these methods perform more poorly and scale such that a large number of training examples is infeasible.  We have included a comparison of these alternative machine learning methods in the Appendices, justifying our decision of using a deep convolutional neural network. One notable limitation of our approach is that the efficient training and evaluation of the \DNN requires uniformity in the input size.  Future work will focus on an approach that would allow transferability to variable input sizes.

Additionally, an electrostatic potential defined on a finite grid can be rotated in integer multiples of \ang{90}, without a change to the electrostatic energies. Convolutional deep neural networks do not natively capture such rotational invariance.  Clearly, this is a problem in any application of deep neural networks (e.g. image classification, etc.), and various techniques are used to compensate for the desired invariance.  The common approach is to train the network on an augmented dataset consisting both of the original training set and rotated copies of the training data \cite{Dieleman2015}.  In this way, the network learns a rotationally invariant set of features.

In demonstration of this technique, we tuned our model trained on the random potentials by training it further on an augmented dataset of rotated random potentials.  We then tested our model on the original testing dataset, as well as a rotated copy of the test set.  The median absolute error in both cases was less than 1.6 mHa.  The median absolute difference in predicted energy between the rotated and unaltered test sets was however larger, at 1.7 mHa.  This approach to training the deep neural network is not absolutely rotationally invariant, however the numerical error experienced due to a rotation was on the same order as the error of the method itself.  Recent proposals to modify the network architecture itself to make it rotationally invariant are promising, as the additional training cost incurred with using an augmented dataset could be avoided \cite{Worrall2016,Dieleman2016}.

In summary, convolutional \DNNs are promising candidates for application to electronic structure calculations as they are designed for data which has a spatial encoding of information.  
\exclude{For this case, even though our \CNN produces a highly accurate result, and does so much faster than our likely less-than-optimal finite-difference numerical solver, the time-to-solution is sufficiently small in absolute terms that the application of a \CNN is not revolutionary.}
As the number of electrons in a system increases, the computational complexity grows polynomially.  Accurate electronic structure methods (e.g. coupled cluster) exhibit a scaling with respect to the number of particles of $N^7$ and even the popular Kohn-Sham formalism of density functional theory scales as $N^3$ \cite{Kohn1995,Kucharski1992}.  The evaluation of a convolutional neural network exhibits no such scaling, and while the training process for more complicated systems would be more expensive, this is a one-time cost.

In this work, we have taken a simple problem (one electron in a confining potential), and demonstrated that a convolutional neural network can automatically extract features and learn the mapping between $V(r)$ and the ground-state energy $\varepsilon_0$ as well as the kinetic energy $\langle\hat T\rangle$, and first excited-state energy $\varepsilon_1$.  Although our focus here has been on a particular type of problem, namely an electron in a confining 2D well, the concepts here are directly applicable to many problems in physics and engineering. Ultimately, we have demonstrated the ability of a \DNN to learn, through example alone, how to rapidly approximate the solution to a set of partial differential equations.  A generalizable, transferable deep learning approach to solving partial differential equations would impact all fields of theoretical physics and mathematics.

\wcexclude{
\section{Acknowledgements}

The authors would like to acknowledge fruitful discussions with P. Bunker, P. Darancet, D. Klug, and  D. Prendergast. K.M. and I.T. acknowledge funding from NSERC and SOSCIP. Compute resources were provided by SOSCIP, Compute Canada, NRC, and an NVIDIA Faculty Hardware Grant.

}



\appendix

\section{Appendices}
\subsection{Appendix A: Comparison of machine learning methods}

One might question the use of a convolutional deep neural network over other more traditional machine learning approaches.  After all, kernel ridge regression, random forests, and artificial neural networks have proven to be quite useful (see main work for references to appropriate work).  Here we compare the use of our convolutional deep neural network approach to kernel ridge regression and random forests, the latter two implemented through Scikit-learn \cite{Pedregosa2012}.

\subsubsection{Kernel ridge regression}

\begin{figure}
  \includegraphics[width=0.89\columnwidth]{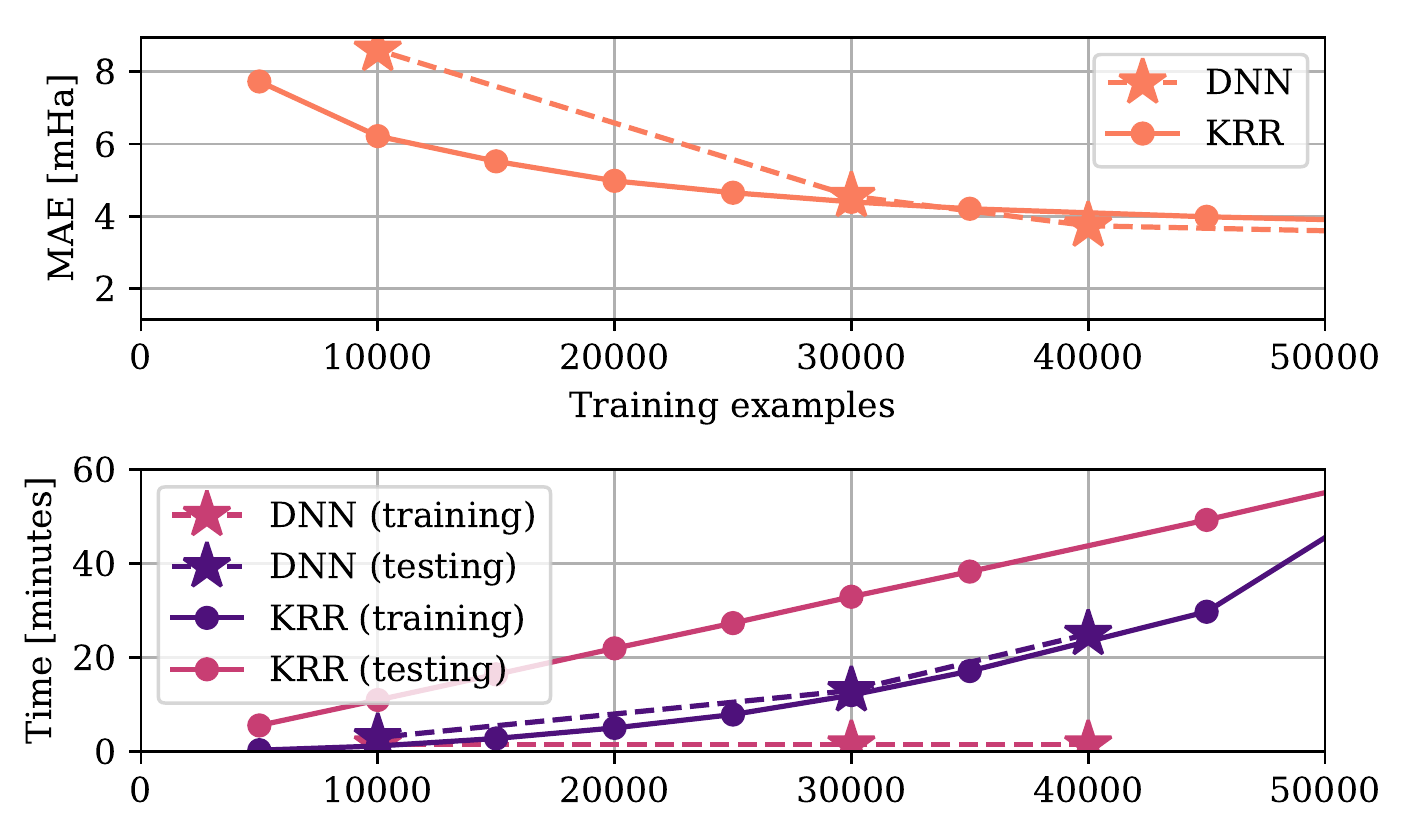}%
  \caption{Kernel ridge regression on simple harmonic oscillator potentials.  When few training examples are provided, kernel ridge regression performs better, however with a larger number of training examples, both methods perform comparably, with DNN slightly better. The training time for kernel ridge regression scales quadratically.  The evaluation time for a fixed number of testing examples scales linearly with respect to the number of \textit{training} examples in the case of kernel ridge regression.  In the case of the deep neural network, the training set size does not affect the testing set evaluation.\label{O_KRR_SHO}}
\end{figure}

\begin{figure}
  \includegraphics[width=0.89\columnwidth]{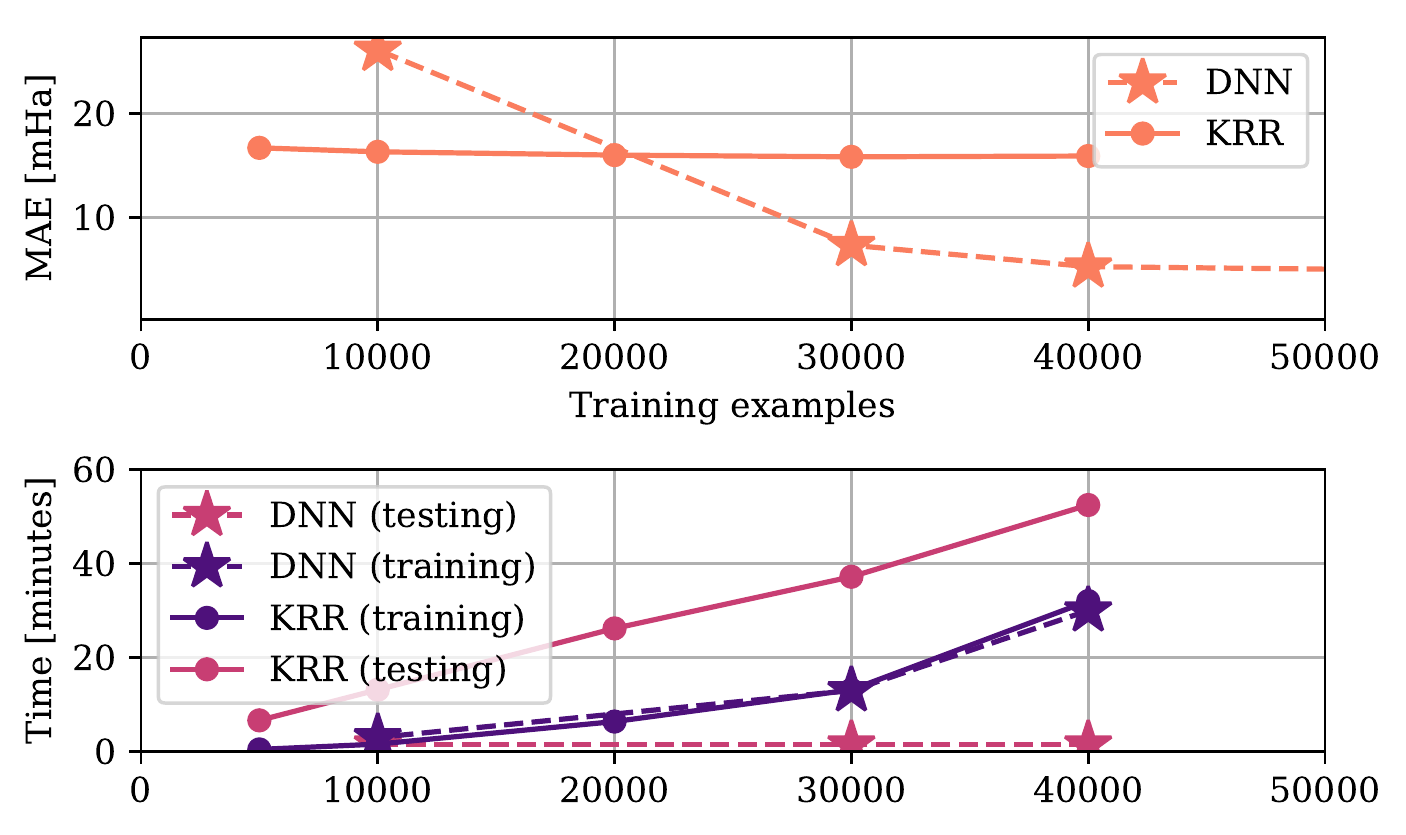}%
  \caption{Kernel ridge regression on random potentials.  When few training examples are present, kernel ridge regression performs better (at constant training time).  This is likely due to the fact that the DNN is only given 10 seconds to run. At larger training set sizes, the deep neural network performs much better; kernel ridge regression barely improves as training set size increases, however training walltime increases dramatically.  \label{O_KRR_RND}}
\end{figure}

We trained a kernel ridge regression model on a training set of simple harmonic oscillator images, recording the walltime (real-world time) taken to train the model.  Then we evaluated the trained model on a test set (the same test set was used throughout).  We recorded both the evaluation walltime and the median absolute error (MAE) observed from the trained model.  We then trained our deep neural network on the same training dataset, allowing it the same training walltime as the KRR model.  We then evaluated the deep neural network on the same testing set of data, again recording the MAE and the evaluation walltime.  This process was repeated for various training set sizes, and on training data from both the simple harmonic oscillator (SHO) and random (RND) datasets.  The results are presented in \figref{O_KRR_SHO} and \figref{O_KRR_RND}.

\subsubsection{Random forests}
\begin{figure}
  \includegraphics[width=0.89\columnwidth]{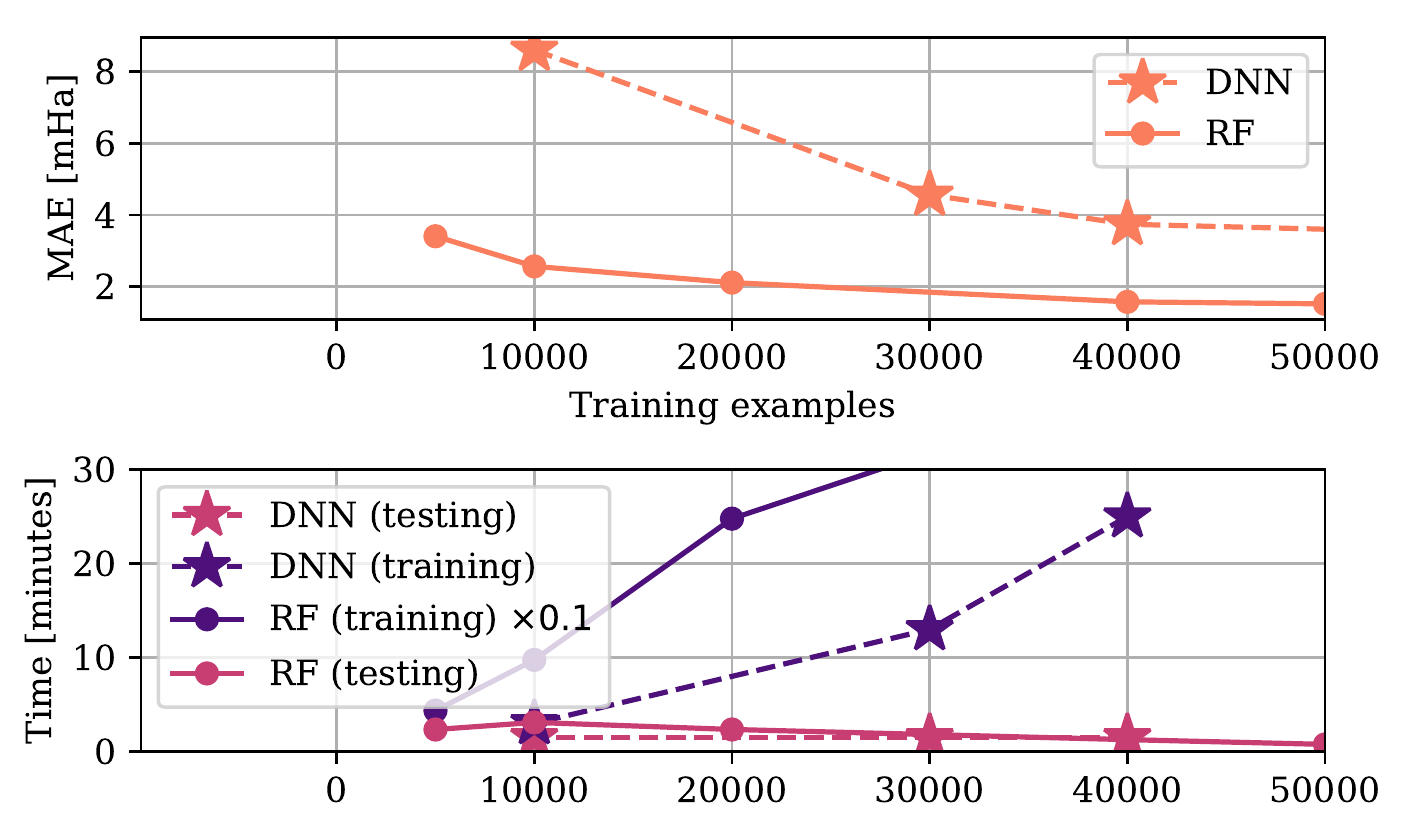}%
  \caption{Random forests and simple harmonic oscillator. Random forests performs better than deep neural networks for all training set sizes on the relatively trivial simple harmonic oscillator dataset.  Random forests takes a very long time to train. Note that the training times plotted above have been scaled by a factor of 0.1 for plotting, and thus the true times are ten times greater than shown. 
  \label{O_RF_SHO}}
\end{figure}

\begin{figure}
  \includegraphics[width=0.89\columnwidth]{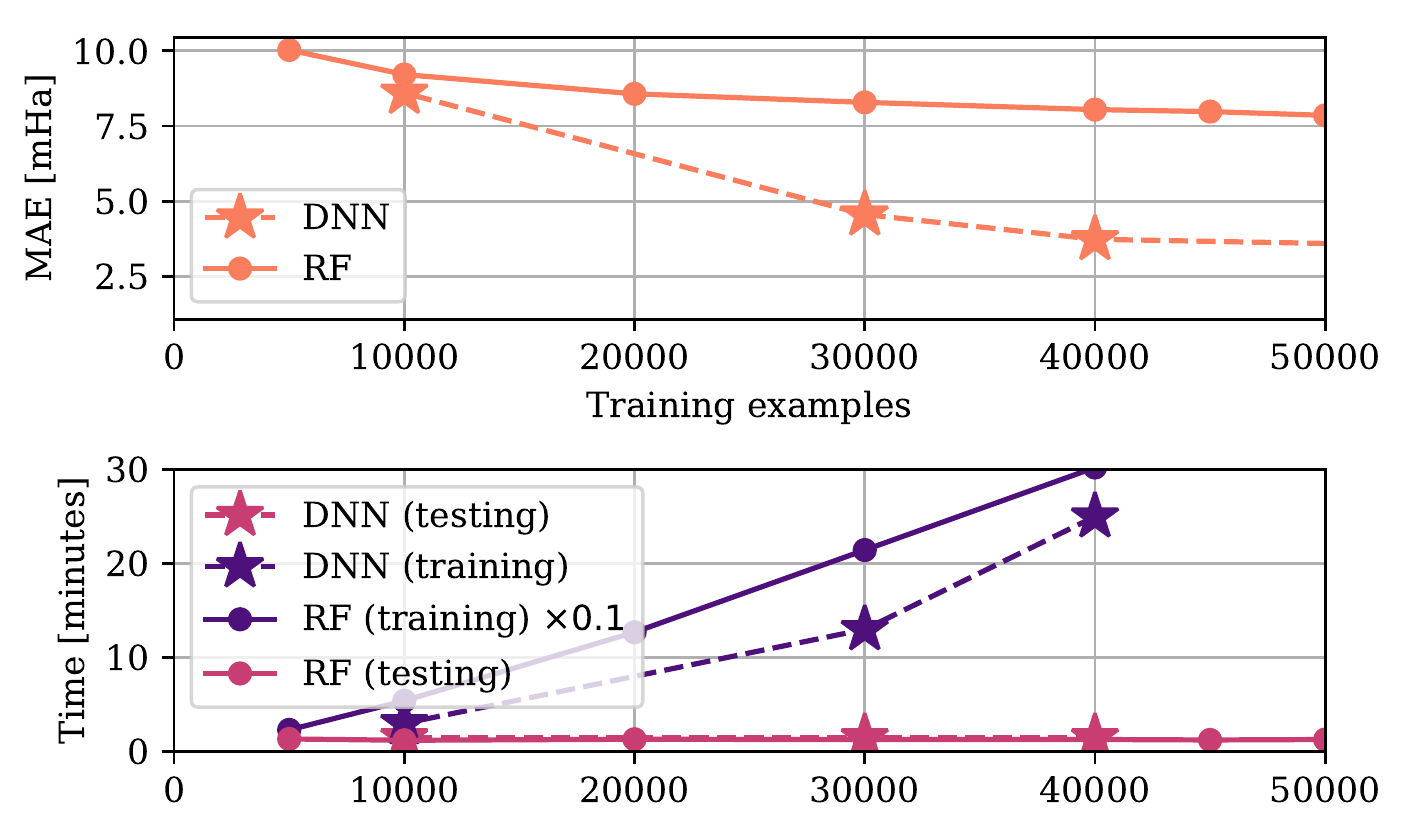}%
  \caption{ Random forests on random potentials.  On the more complicated random potentials, random forests performs significantly worse than the deep neural network.  This combined with the extremely high training time suggests the deep neural network is much better equipped to handle these more varied potentials. \label{O_RF_RND}}
\end{figure}

We carried out an identical process, training a random forests regressor.  The results are presented in \figref{O_RF_SHO} and \figref{O_RF_RND}.

\subsubsection{Discussion}

While the timing comparison is not quantitatively fair (the random forests algorithm is not parallelized and uses only one CPU core. The kernel ridge regression algorithm is parallelized and ran across all available cores, and the deep neural network is highly parallelized via GPU optimization, and runs across thousands of cores), this investigation gives useful insight into the time-to-solution advantages of deep neural networks.  The error rates, however are quantitatively comparable, as the KRR and RF algorithms were permitted to run until convergence. The DNN was able to perform better in most cases given the same amount of walltime.

We see that for all but the simplest cases, our deep neural network is vastly superior to both kernel ridge regression and random forests. For very simple potentials, it is understandable that the machinery of the deep neural network was unnecessary, and that the traditional methods perform well.  For more complicated potentials with more variation in the input data, the deep neural network was able to provide significantly better accuracy in the same amount of time.

\subsection{Appendix B: Dataset generation}

The potentials are defined on a grid from $x,y = -20$ to $20$ a.u. on a $256\times 256$ grid.

\subsubsection{Simple Harmonic Oscillator}

The simple harmonic oscillator (SHO) potentials are generated with with the scalar function

\begin{equation}
V(x,y) =\frac{1}{2} \left(k_x(x - c_x)^2 + k_y(y - c_y)^2\right)
\end{equation}

where $k_x, k_y, c_x$, and $c_y$ are randomly generated according to Table \ref{hoparameterTable}.  The potentials are truncated at 20.0 Ha, (i.e. if $V > 20$, $V=20$).

\begin{table}
 \caption{\label{hoparameterTable}}
 \begin{ruledtabular}
 \begin{tabular}{clcc}
  Parameter & & Lower bound &  Upper bound\\  
  \hline
  $k_x$ & spring constant & 0.0 & 0.16 \\
  $k_y$ & spring constant & 0.0 & 0.16 \\
  $c_x$ & center position & -8.0 & 8.0 \\
  $c_y$ & center position & -8.0 & 8.0 \\ 
 \end{tabular}
 \end{ruledtabular}
\end{table}

\subsubsection{Infinite Well}

The infinite well (IW) potentials are generated with the scalar function 

\begin{equation}
V(x,y) = \begin{cases} 
    0 &  \frac{1}{2}\left(2c_x - L_x\right) < x \le \frac{1}{2}\left(2c_x + L_x\right)\ \text{and}   \\
    & \frac{1}{2}\left(2c_y - L_y\right) < y \le \frac{1}{2}\left(2c_y + L_y\right)  \\
    20 & \text{otherwise}\\
   \end{cases}
\end{equation}
where 20.0 is used as ``numerical infinity'', an appropriate choice given the scale of energies used. Because of the nature of the IW energy, randomly generating $L_x$ and $L_y$ independently leads to a distribution of eneries highly biased toward low energy values (it is more likely to randomly produce a large well than a small).  Since we want a distribution that is as even as possible over the range of energies, we need to take a slightly different approach.  We randomly generate the energy $E$ uniformly on the interval 0 to 0.4 Ha.   We then generate $L_x$ randomly on the interval 4.0 to 15.0, defining the width of the well.  We then solve for the value of $L_y$ that will produce an energy of $E$, given $L_x$, e.g.
\begin{equation}
L_y = 1/\sqrt{ \frac{2E}{\pi^2} - \frac{1}{L_x^2} }
\end{equation}
Not all combinations of $L_x$ and $E$ lead to valid solutions for $L_y$, so we keep trying until one does.  We then swap the values of $L_x$ and $L_y$ with a 50\% probability to prevent one dimension of the well always being larger. This process leads to a relatively even distribution of energies.

\subsubsection{Double-well inverted Gaussians}
The double-well inverted Gaussian (DIG) potentials are generated with the scalar function

\begin{multline}
V(x,y) =  - A_1 \exp \left[ -\left(\frac{x-c_{x_1} }{k_{x_1}} \right) ^2 -\left(\frac{y-c_{y_1} }{k_{y_1}} \right) ^2\right] \\
-  A_2 \exp \left[ -\left(\frac{x-c_{x_2} }{k_{x_2}} \right) ^2 -\left(\frac{y-c_{y_2} }{k_{y_2}} \right) ^2\right]
\end{multline}

where the parameters are randomly sampled from a uniform distribution within the ranges given in Table \ref{igparameterTable}.  These ranges were determined through trial and error to achieve energies in the range of 0 to 400 mHa.

\begin{table}
 \caption{\label{igparameterTable}}
 \begin{ruledtabular}
 \begin{tabular}{clcc}
  Parameter & & Lower bound &  Upper bound\\  
  \hline
  $A_1$ 		& well 1 depth 			& 2.0 & 4.0 \\
  $A_2$ 		& well 2 depth 			& 2.0 & 4.0 \\
  $c_{x_1}$ 	& well 1 center, $x$ 	& -8.0 & 8.0 \\
  $c_{y_1}$ 	& well 1 center, $y$ 	& -8.0 & 8.0 \\
  $c_{x_2}$ 	& well 2 center, $x$ 	& -8.0 & 8.0 \\
  $c_{y_2}$ 	& well 2 center, $y$ 	& -8.0 & 8.0 \\
  $k_{x_1}$ 	& well 1 width 			& 1.6 & 8.0 \\
  $k_{y_1}$ 	& well 1 length 		& 1.6 & 8.0 \\
  $k_{x_2}$ 	& well 2 width 			& 1.6 & 8.0 \\
  $k_{y_2}$ 	& well 2 length 		& 1.6 & 8.0 \\
  \end{tabular}
 \end{ruledtabular}
\end{table}

\subsubsection{Random potentials}

The random potentials are generated through a lengthy process motivated by three requirements: the potentials must
(a) be random (i.e. extremely improbable that two identical potentials ever be generated), (b) be smooth, and (c) go to a maximum of 20.0 at the boundary.

First, we generate a $16\times16$ binary grid of 1s and 0s, and upscale it to $256\times256$.  We then generate a second $16\times 16$ binary grid and and upscale it to $128\times 128$.  We center the smaller grid within the larger grid and then subtract them element-wise. We then apply a Gaussian blur with standard deviation $\sigma_1$, to the resulting image, where $\sigma_1 $ is generated uniformly within the range given in Table \ref{rndparameterTable}. The potential is now random, and smooth, but does not achieve a maximum at the boundary.  

To achieve this, we generate a mask that smoothly goes to zero at the boundary, and 1 in the interior.  We wish the mask to be random, e.g. a randomly generated `blob'.  To generate the blob, we generate $k^2$ random coordinate pairs on a $200\times 200$ grid, where $k$ is an integer between 2 and 7, inclusive.  We then throw away all points that lie inside the covex hull of these points, and smoothly interpolate the remaining points with cubic splines.  We then form a binary mask by filling the inside of this closed blob with 1s, and the outside with 0s.  Resizing the blob to a resolution of $R\times R$, and applying a Gaussian blur with standard deviation $\sigma_2$, we arrive at the final mask.  Here $R$ and $\sigma_2$ are generated uniformly within the ranges given in Table \ref{rndparameterTable}.

Element-wise multiplication of the mask with the random-blurred image gives a random potential that approaches zero at the boundary.  We randomize the ``sharpness'' of the potential by then exponentiating by either $d=0.1, 0.5, 1.0, \text{or} 2.0$, chosen at random with equal probabilities (i.e. $V := V^d$). We then subtract the result from its maximum to invert the well.

\begin{figure}
  \includegraphics[width=0.89\columnwidth]{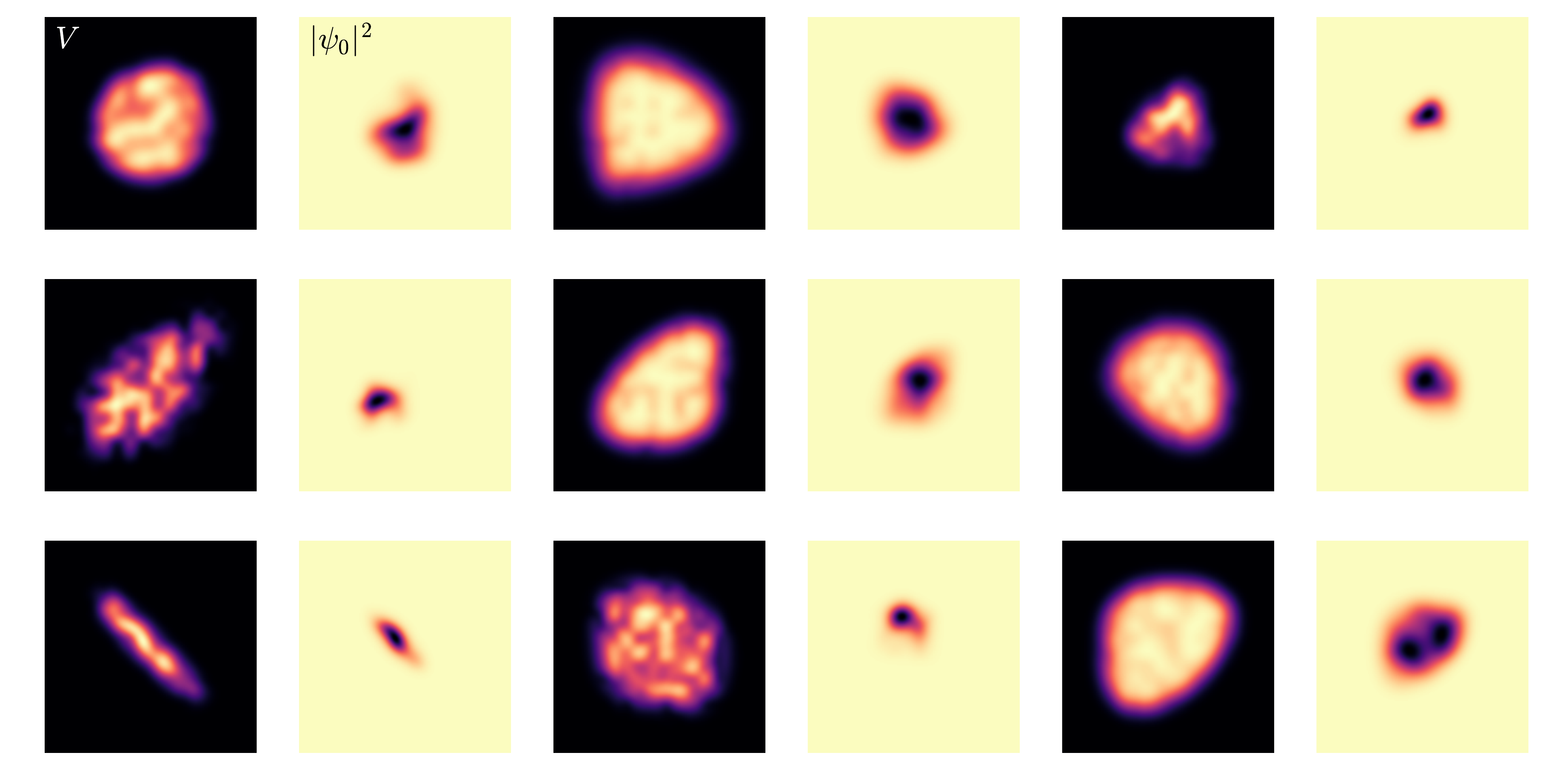}%
  \caption{Some example random potentials $V$ and the norm of their associated ground-state wavefunctions, $|\psi_0|^2$.  \label{wfn_potential_RND}}
\end{figure}

\begin{figure}
  \includegraphics[width=0.89\columnwidth]{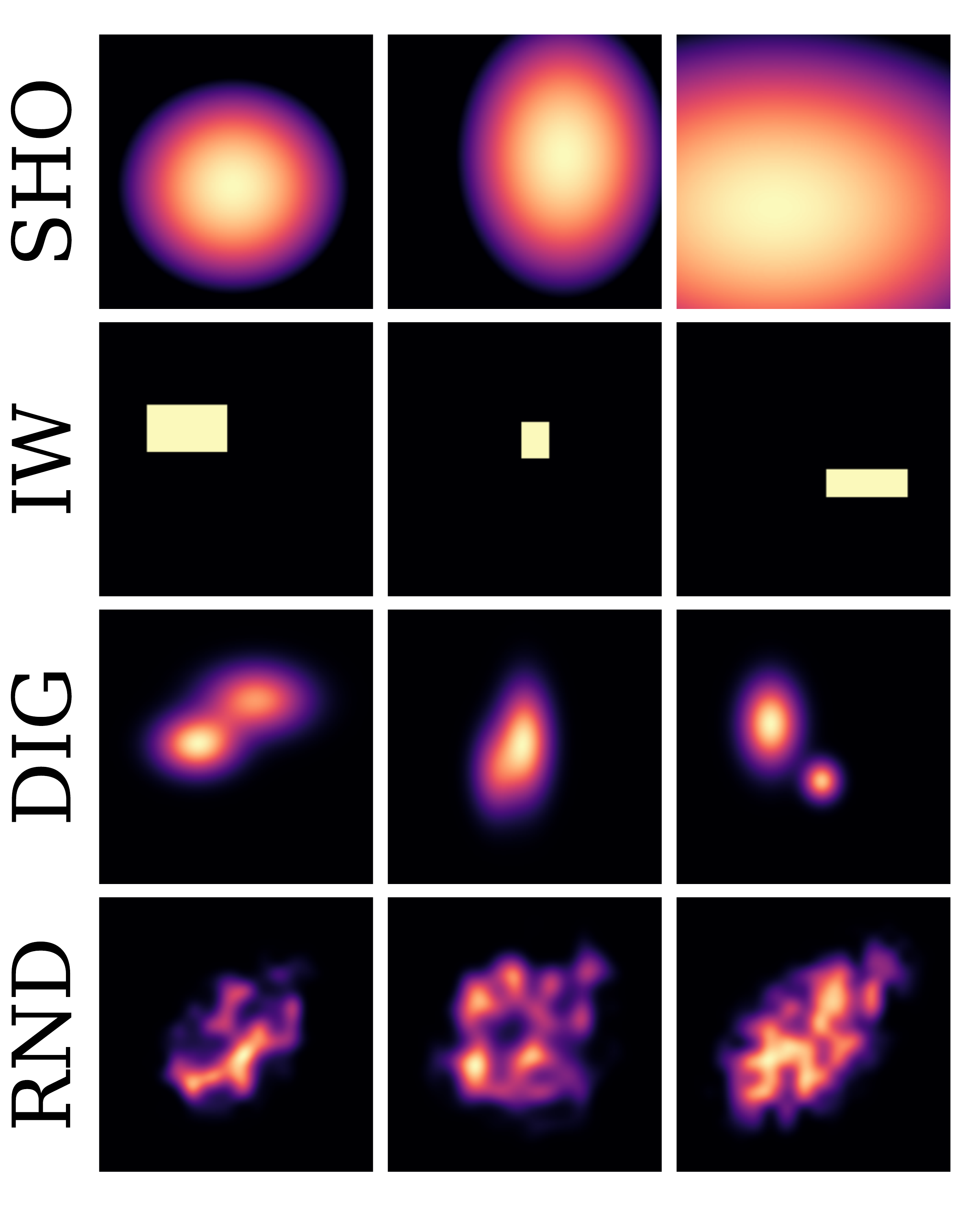}%
  \caption{Examples of the four classes of potentials\label{all_potentials_example}}
\end{figure}

This process, while lengthy, produces very random potentials, of which no two are alike. The energy range of 0 to 400 mHa is appropriate for producing wavefunctions that span a moderate portion of the domain, as seen in \figref{wfn_potential_RND}.
Examples of all classes of potentials can be seen in \figref{all_potentials_example}.

\begin{table}
 \caption{\label{rndparameterTable}}
 \begin{ruledtabular}
 \begin{tabular}{clcc}
  Parameter & & Lower bound &  Upper bound\\  
  \hline
  $\sigma_1$ 		& std. dev.  blur 1   	&    6 		&	 10  \\
  $k$    			& blob points 			&    2   	&    7   \\
  $R$				& blob size    			&	 80		&	 180 \\	
  $\sigma_2$ 		& std. dev. blur 2   	&    10 	&	 16  \\  
  \end{tabular}
 \end{ruledtabular}
\end{table}

\bibliography{../../../../Mendeley/bibtex/MSc-SchrodingerPaper}

\clearpage


\end{document}